\documentclass[preprint,12pt,3p]{elsarticle}
\usepackage[latin9]{inputenc}
\usepackage{amsmath}
\usepackage{amssymb}
\usepackage{graphicx}
\newcommand{\ba}{\begin{eqnarray}}
\newcommand{\ea}{\end{eqnarray}}
\usepackage{amssymb}

\journal{Colloid and Interface Science Communications}

\begin{document}

\begin{frontmatter}

\title{Modeling of a charged dielectric interface: Comparison of the continuum and discrete lattice representations of surface charges}

\author[label1,label2]{Jiaxing Yuan}
\address[label1]{School of Physics and Astronomy, Shanghai Jiao Tong University, Shanghai 200240, China}
\address[label2]{Institute of Natural Sciences, Shanghai Jiao Tong University, Shanghai 200240, China}

\author[label3,label4]{Yanwei Wang\corref{cor1}}
\address[label3]{Department of Chemical and Materials Engineering, School of Engineering, Nazarbayev University, Astana 010000, Republic of Kazakhstan}
\address[label4]{State Key Laboratory of High-Performance Civil Engineering Materials, Jiangsu Sobute New Materials Co., Ltd., Nanjing 211103, China}
\cortext[cor1]{Corresponding author}
\ead{yanwei.wang@nu.edu.kz}

\begin{abstract}
Two main approaches in particle-based simulations for modeling a charged surface are using explicit, discrete charges and continuum, uniform charges. It is well-known that these two approaches could lead to substantially distinct ionic distributions, whereas a systematic exploration of the origin is still absent. In this short communication, we calculate the electrostatic force of a single point charge above a planar substrate characterized by a surface charge density and dielectric mismatch and compare the differences in the electrostatic forces produced by discrete and continuum representations of surface charges. We demonstrate that while the model of uniform surface charges gives a rather simple picture, the model of discrete surface charges can exhibit different scenarios, depending on the respective values of ion-surface distance versus lattice spacing and a self-image interaction parameter.
\end{abstract}



\end{frontmatter}

The structure of electrolytes near charged interfaces is fundamental in controlling stability of colloidal suspension~\cite{linse05,mahynski14}, adsorption of polymers~\cite{messina04,PATTANAIK201841}, and self-assembly of nanoparticles~\cite{walker11,MANDAL20187}. Two main treatments within particle-based simulations for modeling a charged surface are using explicit, discrete charges and continuum, uniform charges. Numerous theories of electrolyte, including the famous Poisson--Boltzmann theory, have been developed on the assumption of uniform surface charge~\cite{andelman95}. However, recent computer simulations~\cite{wang2016examining,wang2017ion} have demonstrated with discrete surface charged groups, one can observe strong charge reversal if the lattice constant becomes notably larger than the counterion diameter, compared to the surfaces with continuum representation. Those results clearly indicate that the discrete and continuum representations of surface charges could lead to substantially different scenarios. However, to our knowledge, a simple understanding of the physics behind such differences is still absent in the literature. In this short communication, we calculate rigorously the electrostatic force on a single point charge above a charged plate and compare the results from discrete and continuum representations of surface charges. By doing so, a simple understanding of the differences between discrete and continuum representations of surface charges is obtained.

\begin{figure}[htb!]
	\centering \includegraphics[width=12cm]{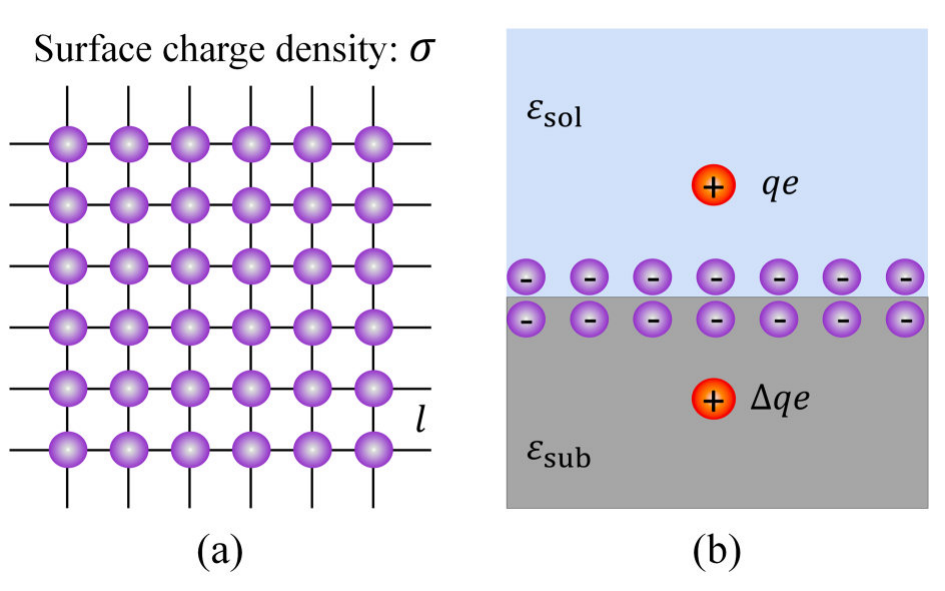}
	\caption{Schematic of a point charge located above a charged, dielectric interfaces. (a) The surface is characterized by a charge density $\sigma$ and a lattice spacing $l$ on an infinite square lattice. (b) The red and purple spheres above the plate represent test charge and surface charges, respectively, for the situation of $\varepsilon_\text{sub}\ll \varepsilon_\text{sol}$. Their image charges with the magnitude weakened by $\triangle$ (see text for definitions) are also sketched.}
	\label{fig:system}
\end{figure}

Consider a charged, planar substrate located at $z=0$ and a test point charge with valence $q$ located at $(\delta_x l,\delta_y l,z)$ where $z>0$, and without loss of generality, $\delta_x\in[0,1/2]$ and $\delta_y\in[0,1/2]$. A schematic of the discrete charged surface is presented in Fig.~\ref{fig:system}. The surface charges are distributed uniformly, forming a two dimensional-square lattice with a lattice constant~$l$. The interface carries a uniform charge density~$\sigma$ such that each surface bead carries a charge of $\sigma{l^2}$. In addition to surface charge, the interface is also characterized by a dielectric contrast which gives rise to a surface polarization charge. The magnitude and sign of the polarization are determined by the dielectric mismatch $\Delta = (\varepsilon_{\mathrm{sol}} - \varepsilon_{\mathrm{sub}}) / (\varepsilon_{\mathrm{sol}} + \varepsilon_{\mathrm{sub}})$ where $\varepsilon_\text{sol}$ and $\varepsilon_\text{sub}$ are the dielectric permittivity of solvent medium and substrate, respectively. $\Delta < 0$ corresponds to a high-permittivity substrate inducing attractive polarization charges; $\Delta > 0$ corresponds to a low-permittivity substrate with repulsive polarization charges, and $\Delta = 0$ corresponds to the case of $\varepsilon_{\mathrm{sol}} = \varepsilon_{\mathrm{sub}}$ in the absence of a dielectric mismatch. Note here we consider the limiting situations of $\Delta=-1$ and $\Delta=+1$, representing metallic substrate and metallic solvent, respectively, to investigate the effects of surface polarization on interfacial electrostatic forces of a point charge.

The electrostatic energy $U_s$ between the test charge and the charged substrate as well as the polarization charge of these surface groups is expressed as
\ba
\beta U_s(z)=\sum _{ { n }_{ x }=-\infty  }^{ \infty  }{ \sum _{ { n }_{ y }=-\infty  }^{ \infty  }{ \frac { (1+\triangle)(2\pi\mu)^{-1}l^2(q\sigma/|q\sigma|) }{ \sqrt { { ({ n }_{ x }l-\delta_xl) }^{ 2 }+{ ({ n }_{ y }l-\delta_yl) }^{ 2 }+{ z }^{ 2 } }  }  }  }  \;,
\ea
where $\beta$ is the Boltzmann factor, $\mu=(2\pi l_B|q\sigma|)^{-1}$ the Gouy-Chapman length~\cite{PHILIPSE201710}, and $l_B$ the Bjerrum length.
From $\beta U_s(z)$, we derive the corresponding electrostatic force $\beta F_s(z)$, which is
\begin{align}
\label{eq:dis}
\beta F_s(z)=-\frac { \partial (\beta U_s) }{ \partial z } =\sum _{ { n }_{ x }=-\infty }^{ \infty  }{ \sum _{ { n }_{ y }=-\infty}^{ \infty  }{ \frac { (1+\triangle)(2\pi\mu)^{-1} (z/l)(q\sigma/|q\sigma|) }{ { (({ { n }_{ x }-\delta_x})^{ 2 }+{ { ({ n }_{ y }-\delta_y)}^{ 2 } }+{ (z/l) }^{ 2 }) }^{ 3/2 } }  }  } \;.
\end{align}
When ${l}\ll z$, $\beta F_s(z)$ can be approximated by a two dimensional integral,
\begin{align}
\label{eq:con}
\nonumber\beta F_s(z)&\simeq\int_{-\infty }^{ +\infty  }\int_{ -\infty }^{ +\infty  }{ dxdy } \frac {(1+\triangle)q{ { \sigma  } }{ l }_{ B }(z/l) }{ { ({ x }^{ 2 }+{ y }^{ 2 }+{ (z/l) }^{ 2 }) }^{ 3/2 } } 
\xrightarrow [ x=rcos(\theta ) ]{ y=rsin(\theta ) } 2\pi\int_{0}^{ +\infty  }{ \frac {(1+\triangle)q{ { \sigma  } }{ l }_{ B }(z/l)rdr }{ { (r^2+{ (z/l) }^{ 2 }) }^{ 3/2 } }  }\\
&=(1+\triangle)\mu^{-1}(q\sigma/|q\sigma|)= \beta F^c_s(z)\;,
\end{align}
where $\beta F^c_s(z)$ represents the force of a point charge above a charged surface with an ``effective'' charge density of $(1+\triangle)\sigma$ in the continuum representation.

\begin{figure}[htb!]
	\centering \includegraphics[width=12cm]{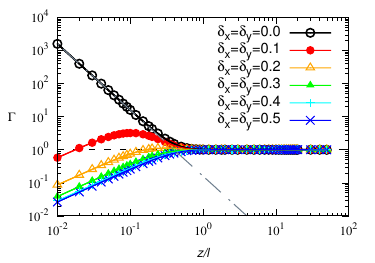}
	\caption{Numerical results of $\Gamma$ at various $z/l$, $\delta_x$ and $\delta_y$ values. The dashed line corresponds to the asymptotic behavior at $z/l \gg 1$ where $\Gamma = 1$, and the dot-dashed line corresponds to the asymptotic behavior at $z/l \ll 1$ and $\delta_x = \delta_y = 0$ (see text for details).}
	\label{fig:Fig2}
\end{figure}

To compare the differences of electrostatic forces in discrete and continuum model, we calculate numerically the ratio of $F_s(z)$ to $F^c_s$, 
\ba
\label{eq:ratio}
\Gamma=\frac{F_s}{F^c_s}=\sum _{ { n }_{ x }=-\infty }^{ \infty  }{ \sum _{ { n }_{ y }=-\infty}^{ \infty  }{ \frac { (2\pi)^{-1} (z/l) }{ { (({ { n }_{ x }-\delta_x})^{ 2 }+{ { ({ n }_{ y }-\delta_y)}^{ 2 } }+{ (z/l) }^{ 2 }) }^{ 3/2 } }  }  } \;.
\ea
Fig.~\ref{fig:Fig2} shows our results of $\Gamma$ as a function of $z/l$ for various $\delta_x$ and $\delta_y$. Some interesting features are revealed for the domain where $z/l \ll 1$. When $\delta_x=\delta_y=0$, $\Gamma$ is dominant by the $n_x=n_y=0$ term, and diverges as $(2\pi)^{-1}(z/l)^{-2}$.

Clearly, as $z/l \ll 1$, the electrostatic force of a point charge near the discretely charged interface can become significantly larger compared to that from the continuum model, indicating discretely charged pattern can enhance the ion-surface correlation. This provides an intuitive understanding of the so-called ``giant charge reversal'' observed in simulations\cite{wang2016examining,wang2017ion}.

When $\delta_x$ and $\delta_y$ are nonzero, the numerical data (Fig.~\ref{fig:Fig2}) suggests that the lateral ion distribution is non-uniform in the discrete surface-charge model. For discretely charged surfaces with a large lattice spacing, the surface counterions are localized and binding around individual charged beads, as confirmed by simulations~\cite{wang2016examining,calero2010interaction}. We also note the connection to the recent experimental observation where multivalent ions induce lateral structural inhomogeneities in polyelectrolyte brushes near surfaces~\cite{yu17}.

\begin{figure}[htb!]
	\centering \includegraphics[width=12cm]{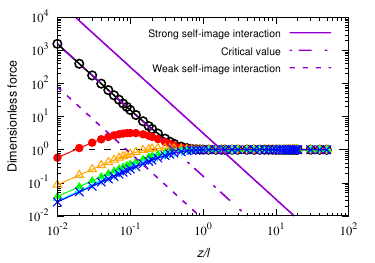}
	\caption{Comparing the self-image interaction at different strengths with the ion-surface charge interaction.}
	\label{fig:Fig3}
\end{figure}

Apart from the ion-surface charge interactions, the point charge also interacts with its induced-surface polarization (represented via an image charge), and the corresponding force $\beta F_i$ is expressed as
\ba
\beta F_i(z)=\triangle\frac{q^2l_B}{4z^2}\;.
\ea
To compare the relative magnitude between ion-surface interaction and self-image interaction, in Fig.~\ref{fig:Fig3} we plot the absolute dimensionless force $|F_i/F^c_s|$ in comparison with $F_s/F^c_s$. In other words, we choose to use $|F^c_s|$ as the characteristic force scale. It follows that
\ba
|\frac{F_i}{F^c_s}|=\frac{|\triangle|}{1+\triangle}\frac{q^2l_B\mu}{4z^2}=\frac{|\triangle|}{1+\triangle}\frac{q^2l_B\mu}{4l^2}(z/l)^{-2} = K(z/l)^{-2} \;,
\ea
where the dimensionless number $K$ is introduced to denote the strength of self-image interactions, and is defined as
\ba
K =\frac{|\triangle|}{1+\triangle}\frac{q^2l_B\mu}{4l^2} \;.
\ea
Compared to the asymptotic behavior of Eq.~(\ref{eq:ratio}) when $\delta_x=\delta_y=0$, we find a critical coupling of $K^{*} = (2\pi)^{-1}$. For strong coupling $K \gg K^{*}$, the self-image interaction dominates the ion-surface interaction for a wide range of $z/l$, whereas for weak coupling $K \ll K^{*}$, the relative strength between self-image interaction and ion-surface charge interaction depends on the position $z/l$. Obviously, at the far field $z/l\gg 1$, the ion-surface interaction always plays the dominant role.

Taking into account the overall interaction between a point charge and a charged dielectric interface, Fig.~\ref{fig:Fig3} shows that while the continuum, uniform surface charge model gives a rather simple picture, the discrete surface charge model can offer a range of different scenarios (even for such a toy model!), depending on the values of $z/l$ and $K$. The finite size of the surface ionic groups and that of the counterions sets a lower bound of $z$ close to the surface. Let $a$, typically $5\sim7 \, \mathring { A } $, be the diameter of the surface ionic groups and the counterions. Our results demonstrate that the interface with discretely distributed surface charges can lead to highly enhanced ion-surface binding and non-uniform lateral ion distribution (relevant to ion transport near interfaces) for $l\gg 5\sim7 \, \mathring { A } $ (large lattice spacing), in particular when $K \ll K^{*}$ (weak self-image coupling). We expect these results deliver a simple, yet fundamental insight into understandings of ion-surface electrostatic interactions near a charged, dielectric interface.

\section*{Acknowledgements}
Y. Wang acknowledges the support of the National Key Research and Development Program of China (2017YFB0310100) and the State Key Laboratory of High-Performance Civil Engineering Materials (2014CEM001).

 \bibliographystyle{elsarticle-num-names}
\bibliography{manuscript}

\begin{thebibliography}{12}
\expandafter\ifx\csname natexlab\endcsname\relax\def\natexlab#1{#1}\fi
\providecommand{\url}[1]{\texttt{#1}}
\providecommand{\href}[2]{#2}
\providecommand{\path}[1]{#1}
\providecommand{\DOIprefix}{doi:}
\providecommand{\ArXivprefix}{arXiv:}
\providecommand{\URLprefix}{URL: }
\providecommand{\Pubmedprefix}{pmid:}
\providecommand{\doi}[1]{\href{http://dx.doi.org/#1}{\path{#1}}}
\providecommand{\Pubmed}[1]{\href{pmid:#1}{\path{#1}}}
\providecommand{\bibinfo}[2]{#2}
\ifx\xfnm\relax \def\xfnm[#1]{\unskip,\space#1}\fi
\bibitem[{Linse(2005)}]{linse05}
\bibinfo{author}{P.~Linse},
\newblock \bibinfo{title}{Simulation of charged colloids in solution},
\newblock in: \bibinfo{booktitle}{Advanced Computer Simulation Approaches for
  Soft Matter Sciences II (Advances in Polymer Science)}, volume
  \bibinfo{volume}{185}, \bibinfo{publisher}{Springer}, \bibinfo{year}{2005},
  pp. \bibinfo{pages}{111--162}.
\bibitem[{Mahynski et~al.(2014)Mahynski, Panagiotopoulos, Meng, and
  Kumar}]{mahynski14}
\bibinfo{author}{N.~A. Mahynski}, \bibinfo{author}{A.~Z. Panagiotopoulos},
  \bibinfo{author}{D.~Meng}, \bibinfo{author}{S.~K. Kumar},
\newblock \bibinfo{title}{Stabilizing colloidal crystals by leveraging void
  distributions},
\newblock \bibinfo{journal}{Nat. Commun.} \bibinfo{volume}{5}
  (\bibinfo{year}{2014}) \bibinfo{pages}{4472}.
\bibitem[{Messina(2004)}]{messina04}
\bibinfo{author}{R.~Messina},
\newblock \bibinfo{title}{Effect of image forces on polyelectrolyte adsorption
  at a charged surface},
\newblock \bibinfo{journal}{Phys. Rev. E} \bibinfo{volume}{70}
  (\bibinfo{year}{2004}) \bibinfo{pages}{051802}.
\bibitem[{Pattanaik and Venugopal(2018)}]{PATTANAIK201841}
\bibinfo{author}{A.~Pattanaik}, \bibinfo{author}{R.~Venugopal},
\newblock \bibinfo{title}{Investigation of adsorption mechanism of reagents
  (surfactants) system and its applicability in iron ore flotation--an
  overview},
\newblock \bibinfo{journal}{Colloid Interface Sci. Commun.}
  \bibinfo{volume}{25} (\bibinfo{year}{2018}) \bibinfo{pages}{41--65}.
\bibitem[{Walker et~al.(2011)Walker, Kowalczyk, Olvera~de La~Cruz, and
  Grzybowski}]{walker11}
\bibinfo{author}{D.~A. Walker}, \bibinfo{author}{B.~Kowalczyk},
  \bibinfo{author}{M.~Olvera~de La~Cruz}, \bibinfo{author}{B.~A. Grzybowski},
\newblock \bibinfo{title}{Electrostatics at the nanoscale},
\newblock \bibinfo{journal}{Nanoscale} \bibinfo{volume}{3}
  (\bibinfo{year}{2011}) \bibinfo{pages}{1316--1344}.
\bibitem[{Mandal et~al.(2018)Mandal, Konduru, Ramazani, Molina, and
  Larson}]{MANDAL20187}
\bibinfo{author}{T.~Mandal}, \bibinfo{author}{N.~V. Konduru},
  \bibinfo{author}{A.~Ramazani}, \bibinfo{author}{R.~M. Molina},
  \bibinfo{author}{R.~G. Larson},
\newblock \bibinfo{title}{Effect of surface charge and hydrophobicity on
  phospholipid-nanoparticle corona formation: A molecular dynamics simulation
  study},
\newblock \bibinfo{journal}{Colloid Interface Sci. Commun.}
  \bibinfo{volume}{25} (\bibinfo{year}{2018}) \bibinfo{pages}{7--11}.
\bibitem[{Andelman(1995)}]{andelman95}
\bibinfo{author}{D.~Andelman},
\newblock \bibinfo{title}{Electrostatic properties of membranes: the
  poisson-boltzmann theory},
\newblock in: \bibinfo{booktitle}{Handbook of biological physics},
  volume~\bibinfo{volume}{1}, \bibinfo{publisher}{Elsevier},
  \bibinfo{year}{1995}, pp. \bibinfo{pages}{603--642}.
\bibitem[{Wang and Ma(2016)}]{wang2016examining}
\bibinfo{author}{Z.-Y. Wang}, \bibinfo{author}{Z.~Ma},
\newblock \bibinfo{title}{Examining the contributions of image-charge forces to
  charge reversal: Discrete versus continuum modeling of surface charges},
\newblock \bibinfo{journal}{J. Chem. Theory Comput.} \bibinfo{volume}{12}
  (\bibinfo{year}{2016}) \bibinfo{pages}{2880--2888}.
\bibitem[{Wang and Wu(2017)}]{wang2017ion}
\bibinfo{author}{Z.-Y. Wang}, \bibinfo{author}{J.~Wu},
\newblock \bibinfo{title}{Ion association at discretely-charged dielectric
  interfaces: Giant charge inversion},
\newblock \bibinfo{journal}{J. Chem. Phys.} \bibinfo{volume}{147}
  (\bibinfo{year}{2017}) \bibinfo{pages}{024703}.
\bibitem[{Philipse et~al.(2017)Philipse, Tuinier, Kuipers, Vrij, and
  Vis}]{PHILIPSE201710}
\bibinfo{author}{A.~P. Philipse}, \bibinfo{author}{R.~Tuinier},
  \bibinfo{author}{B.~Kuipers}, \bibinfo{author}{A.~Vrij},
  \bibinfo{author}{M.~Vis},
\newblock \bibinfo{title}{On the repulsive interaction between strongly
  overlapping double layers of charge-regulated surfaces},
\newblock \bibinfo{journal}{Colloid Interface Sci. Commun.}
  \bibinfo{volume}{21} (\bibinfo{year}{2017}) \bibinfo{pages}{10--14}.
\bibitem[{Calero and Faraudo(2010)}]{calero2010interaction}
\bibinfo{author}{C.~Calero}, \bibinfo{author}{J.~Faraudo},
\newblock \bibinfo{title}{The interaction between electrolyte and surfaces
  decorated with charged groups: A molecular dynamics simulation study},
\newblock \bibinfo{journal}{J. Chem. Phys.} \bibinfo{volume}{132}
  (\bibinfo{year}{2010}) \bibinfo{pages}{024704}.
\bibitem[{Yu et~al.(2017)Yu, Jackson, Xu, Brettmann, Ruths, de~Pablo, and
  Tirrell}]{yu17}
\bibinfo{author}{J.~Yu}, \bibinfo{author}{N.~E. Jackson},
  \bibinfo{author}{X.~Xu}, \bibinfo{author}{B.~K. Brettmann},
  \bibinfo{author}{M.~Ruths}, \bibinfo{author}{J.~J. de~Pablo},
  \bibinfo{author}{M.~Tirrell},
\newblock \bibinfo{title}{Multivalent ions induce lateral structural
  inhomogeneities in polyelectrolyte brushes},
\newblock \bibinfo{journal}{Sci. Adv.} \bibinfo{volume}{3}
  (\bibinfo{year}{2017}) \bibinfo{pages}{eaao1497}.

\end{thebibliography}

\end{document}